# A Quantum Past
## Jeremy Bernstein

I deduce two general conclusions from these thought-experiments. First, statements about the past cannot in general be made in quantum-mechanical language. We can describe a uranium nucleus by a wave-function including an outgoing alpha-particle wave which determines the probability that the nucleus will decay tomorrow. But we cannot describe by means of a wave-function the statement, ``This nucleus decayed yesterday at 9 a.m. Greenwich time''. As a general rule, knowledge about the past can only be expressed in classical terms. My second general conclusion is that the ``role of the observer'' in quantum mechanics is solely to make the distinction between past and future. The role of the observer is not to cause an abrupt ``reduction of the wave-packet'', with the state of the system jumping discontinuously at the instant when it is observed. This picture of the observer interrupting the course of natural events is unnecessary and misleading. What really happens is that the quantum-mechanical description of an event ceases to be meaningful as the observer changes the point of reference from before the event to after it. We do not need a human observer to make quantum mechanics work. All we need is a point of reference, to separate past from future, to separate what has happened from what may happen, to separate facts from probabilities. Freeman Dyson[1]

A long time ago Murray Gell-Mann told me that Feynman had once said to him that quantum mechanics could not account for history. I think that this is another way of stating Dyson's point. The first thing I want to do in this essay is to explain my understanding of what this means. I am in these matters very conscious of my favorite aphorism of Niels Bohr. He cautioned us not to speak-or write-more clearly than we think. In any event the first question to be answered is what is understood by "quantum mechanics." In this I shall frequently employ John Bell's acronym FAPP-For All Practical Purposes. The quantum mechanics I shall refer to here is FAPP mechanics. It is the quantum mechanics say in Dirac's book. If you haven't read his book for awhile, I suggest you look at the opening chapter. There is not one word about any of the foundational issues that are now fashionable. You will find no reference to the Einstein, Podolsky, Rosen paper. The word "complementarity" is not found, nor is "entanglement", nor "non-locality." There is no measurement "problem" but only a matter of fact statement of what a measurement is. John Bell once told me that Dirac had said to a colleague that it was a good book, but that the first chapter was missing. As far as his book is concerned, his only interest is how to formulate the theory in order to solve problems. In short, it is pure FAPP.

---

[1] Thought Experiments in Honor of John Archibald Wheeler. In Science and Ultimate Reality, Cambridge University Press, New York, 2004, p.89

In FAPP language we have a quantum mechanical system described by a wave function ψ(t), I am only interested in the time variable. The wave function obeys a Schrödinger equation with a Hamiltonian H. The formal solution to this equation is ψ(t)=exp(iHt)ψ(0). Throughout I am setting ℏ=1. Thus to recover Ψ(0) from ψ(t) all we have to do is to multiply by exp(-iHt). Haven't we then recovered the past? What is all the fuss about? The problem is that there is more to life than the wave function. There are the "observables" which represent what we really want to know about the system. These observables are described by Hermitian operators A.B.C and so on. We can expand ψ in a sum over the ortho-normal eigen -functions of any of these operators. The coefficients in the expansion are related to the probabilities that in a measurement the system will be found to have one of these eigen-values. This is "Born's rule" and in FAPP it must be assumed.

To find which of these eigen-values the system actually has, we must perform a measurement. Stripped to its essence the apparatus that produces this measurement projects out from the sum of eigen-functions one of them. After the measurement the rest of the terms in the sum disappear. Using the term of art, the wave function "collapses." It is at this point that we lose our capacity to reconstruct the past. Projection operators are singular. They do not have inverses. All the king's horses and all the king's men cannot put the wave function back together again. It was von Neumann in the early 1930's who first noted that in FAPP mechanics there were two kinds of processes. There were processes that could be described by a Schrödinger equation and there were measurements which could not. He did not, as far as I know, comment on what this implied for retro diction. A case in point is an electron described by a spherically symmetric Schrödinger wave. If this electron strikes a detector is does so at a place-a spot. After this happens all trace of the spherically symmetric wave function vanishes.

I have certainly not made a careful search of the literature but among the founding fathers of FAPP I can come up with only two references that deal with the matter of the quantum past. One is Heisenberg and the other is a paper by Einstein, Richard Tolman, and Boris Podolsky, "Knowledge of Past and Future in Quantum Mechanics" which they wrote in 1931 when Einstein was spending time at CalTech. I think that this was Einstein's first paper in English.[2] First, Heisenberg.

---

[2] Phys. Rev. 37 (1931) 78.

In 1929 he gave a series of lectures on the quantum theory at the University of Chicago. These were published in 1930 in a book entitled The Physical Principles of the Quantum Theory.[3] There is one paragraph devoted to the quantum past which I will quote in its entirety. [4]

"The uncertainty principle refers to the degree of indeterminateness in the possible present knowledge of the simultaneous values of various quantities with which the quantum theory deals; it does not restrict, for example, the exactness of a position measurement alone or a velocity measurement alone. Thus suppose that the velocity of a free electron is precisely known, while the position is completely unknown. Then the principle states that every subsequent observation of the position will alter the momentum by an unknown and undeterminable amount such that after carrying out the experiment our knowledge of the electronic motion is restricted by the uncertainty relation. This may be expressed in concise and general terms by saying that every experiment destroys some of the knowledge of the system which was obtained by previous experiments."

Then he writes,

---

[3] Dover Press, New York , 1930.
[4] Heisenberg op. cit. p.20

"This formulation makes it clear that the uncertainty relation does not refer to the past: if the velocity of the electron is at first known and the position then exactly measured the position for times previous to the measurement may be calculated. Thus for the past times ΔxΔp is smaller than the usual limiting value, but this knowledge of the past is of a purely speculative character, since it can never (because of the unknown change in momentum caused by the position measurement) be used as an initial condition in any calculation of the future progress of the electron and thus cannot be subjected to experimental verification. It is a matter of personal belief whether such a calculation concerning the past history of the electron can be ascribed any physical reality or not."

Clearly the electron had a past history but what Heisenberg seems to be saying is that it cannot be described by quantum mechanics.

The Einstein,Tolman, Podolsky paper makes a somewhat different point. They produce a thought experiment which I will describe in which two particles, electrons say, move along two trajectories one long and one short to a common detector. They argue that if retro diction were allowed for the particle that arrives first then you could make a prediction about the arrival of the second particle that would violate the uncertainty principle. The conclusion is that the uncertainty principle applies to reconstructions of the past as well as predictions of the future. On its face this seems to be inconsistent with everything we believe about the past. I may conjecture that the Sun will come up tomorrow, but I **know** that the Sun came up yesterday. Here is how they begin their brief paper.

"It is well-known that the principles of quantum mechanics limit the possibilities of exact predictions as to the future path of a particle. It has sometimes been supposed, nevertheless, that the quantum mechanics would permit an exact description of the past path of a particle"

One would like to know who "supposed" this. There are no references of any kind in their note. They go on,

"The purpose of the present note is to discuss a simple ideal experiment which shows that the possibility of describing the past path of one particle would lead to predictions as to the future behaviour [sic] of a second particle of a kind not allowed in the quantum mechanics. It will hence be concluded that the principles of quantum mechanics actually involve an uncertainty in the description of past events which is analogous to the uncertainty in the prediction of future events. And it will be shown for the case in hand, that this uncertainty in the description of the past arises from a limitation of the knowledge that can be obtained by measurement of momentum."[5]

In this setup the authors imagine a box on some sort of scale. Inside the box are particles in agitation. There are two holes and a shutter that opens and closes them. Einstein buffs will be reminded of a similar apparatus that Einstein introduced in the 1930 Solvay meeting. This one he used to "refute" the Heisenberg energy time uncertainly relation. It will be recalled that Bohr pointed out that Einstein had left out the gravitational time variation of the clock in the box as it changed positions in the gravitational field of the Earth. Once this was taken into account the uncertainty principle was saved. In their paper the authors insist that the clock of the observer at the end of the shorter path is far enough away so that it will not be perturbed by any gravitational effects due to the weighing of the box. Once bitten twice shy. The shutter is opened briefly and out fly two particles one of which takes the short route and one of which takes the long route to the detector. The box is weighed before and after the particles are released and thus the total energy of the two particles is known.

---

[5] Einstein at al, op cit p780.

The observer at the detector measures the momentum of the particle arriving by the short route and its time of arrival. We now know the energy, and the speed of the first particle. We also know how far it has gone since we have measured that before hand. Thus we can apparently say at what time the shutter opened and how much energy the second particle has. Hence we might argue that since we know its speed and the distance it has to travel we can say exactly at what time it will arrive at the detector and with what energy thus violating the uncertainty principle. The flaw in all of this, as the authors point out, is the assumption, reasonable on classical grounds, that we can determine the momentum of the first particle along the trajectory by a retrospective argument. If we want to be consistent quantum mechanically we must say that the particle has no momentum until we measure it and that once we measure it its future momentum is uncertain. The authors conclude,

"It is hence to be concluded that the principles of quantum mechanics must involve an uncertainty in the description of past events which is analogous to the uncertainty in the prediction of future events. It is also to be noted that although it is possible to measure the momentum of a particle and follow this with a measurement of position , this will not give sufficient information for a complete reconstruction of its past path, since it has been shown that there can be no method for measuring the momentum of a particle without changing its value…"[6]

What I find remarkable is that two years later Einstein was in Princeton working with Podolsky and Rosen on the inability of the quantum theory ,as they saw it, to include all elements of reality. The elements of reality that they say are not included seem to me small beer as compared to the entire past! *Eppur si muove*. The Earth does move. There is a past. Hitler **is** dead. There **was** a total eclipse of the Sun on May 14, 1230. Yet FAPP mechanics cannot describe this without uncertainties. On this ,as far as I can see, Einstein was silent.

There are various attitudes we can adopt towards this. The FAPP attitude was well-summarized by Alfred E, Newman. "What me worry?" After all, as scientists we are concerned with making predictions. Leave the retro dictions to the historians. Of course there is a good deal of scientific enterprise devoted to using quantum mechanics to estimate things like the amount of helium produced in the first three minutes after the Big Bang. That is retro-diction big time. We could simply accept the fact that according to FAPP mechanics the uncertainty principle applies to the past as well as the future. After all in principle the uncertainty principle affects everything we do, I am sure that it affects the trajectory of my bicycle commute, but FAPP, that is the least of my worries. However accepting this does not deal with the quantum measurement issue. After such a measurement FAPP theory tells us that part-indeed most- of the wave function disappears along with our knowledge of the past. That is something to think about. It is clear that if we take this seriously we have to go beyond FAPP.

---

[6] Einstein et al op cit, p.781.

It seems to me that any interpretation of the quantum theory that addresses this must have the feature that measurements are simply just another interaction like the rest. Von Neumann's notion that there were two classes of interactions one whose time evolution could be described by a Schrödinger equation and one of which couldn't, has to be abandoned. I will discuss two proposals for doing this each of which has its adherents and its detractors. On the one hand I am going to discuss what I will call "Bohmian mechanics" a term which David Bohm, who invented this approach , apparently did not like. As far as he was concerned, he was just doing quantum mechanics but in a different way. However nearly everyone else calls it Bohmian mechanics-so will I. On the other hand, I am going to discuss the "decoherent history" interpretation which Murray Gell-Mann and Jim Hartle have done the most on. Sometimes this is called the "many worlds" interpretation, but not by them. I think that the term "many worlds" is misleading. As far as we know there is one world, the one we live in. First Bohmian mechanics.

To simplify things I will restrict myself to a single particle interacted on by an external force with a potential V(**x**). In Bohmian mechanics this particle has a real classical position **X**(t). There are no uncertainties here. The particle follows a classical trajectory which is determined by a first order differential equation that I will write down. As we shall see, what drives this differential equation is a wave function ψ(**x**,t) where **x** is any point In space including **X**. ψ satisfies the Schrödinger equation

$$i\partial/\partial t \psi(\mathbf{x},t) = H\psi(\mathbf{x},t).$$

Here H is the Hamiltonian that includes the potential V(**x**). To write the equation for **X**(t) we introduce the current **J**(x,t)

$$J(\mathbf{x},t) = \frac{1}{2im}(\psi^*(\mathbf{x},t)\partial\psi(\mathbf{x},t) - \psi(\mathbf{x},t)\partial\psi^*(\mathbf{x},t))$$

where m is the mass of the particle. We also introduce the density $\rho(\mathbf{x},t)$ where

$$\rho(\mathbf{x},t) = \psi^*(\mathbf{x},t)\psi(\mathbf{x},t).$$

Using the Schrödinger equation on can establish the continuity equation

$$\partial/\partial t \psi + \partial \cdot \mathbf{J} = 0.$$

The equation for the trajectory of $\mathbf{X}(t)$ is given by-an assumption

$$d\mathbf{X}(t)/dt = \mathbf{J}(\mathbf{X}(t).t)/\rho(\mathbf{X}(t),t)$$

It is comforting to report that for a free particle, V=O,

$$d\mathbf{X}(t)/dt = \mathbf{p}/m.$$

Incidentally, Bohmian mechanics is very often called a "hidden variable" theory. It seems to me that this is a misnomer. There is nothing hidden about the position variables of the particles. It would I think be better to call it a "classical variable" theory.

It is not my purpose here to give many examples of Bohmian mechanics in action. But let me describe one that I think is very impressive. That is the Bohmian analysis of the two-slit experiment. I think that we all remember that when we were first told about it the question that immediately came to mind was how does the electron "know" that the other slit is open or closed? Does the electron somehow go through both slits at once? We are FAPPed into submission by being told that this is a question that FAPP mechanics cannot and will not answer. If we want to observe the electron at one slit we will destroy the interference pattern. Bohmian mechanics does away with all this FAPPness. See the diagram below.

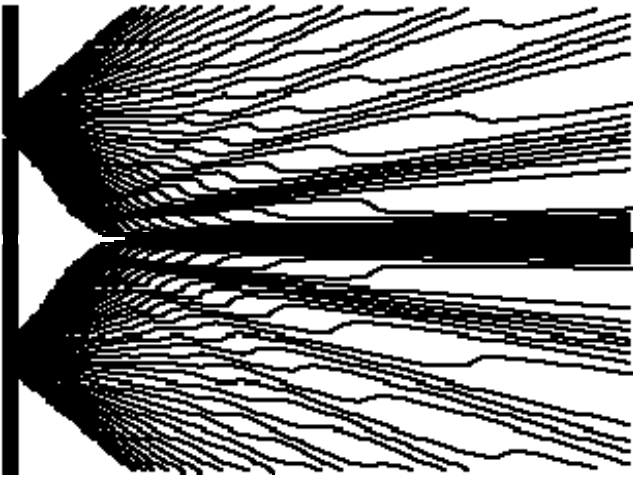

The lines represent the different possible electron trajectories after the electron has passed through one of the slits. The electron goes through only one slit while the guide wave goes through both. This is why single electrons can produce the diffraction pattern when they are let through the slits one at a time. There is no collapse of the wave function or anything like it. Likewise in a measurement there is no collapse of the wave function. For example, suppose you have an observable with two possible values. Then depending on the setting of the detector the particle will be guided to one trajectory or the other. After registering its arrival the particle can continue on. The wave function has not collapsed but the part appropriate to the other possible trajectory "decoheres"-separates out-and can no longer influence the path of the particle.

All of this can be run backwards in time to re-create history. One may wonder where in all of this are the uncertainties of the quantum theory. To get a feel for what is going on, let's consider the S-wave particle I discussed earlier. The S-wave is symmetrically spread over space. But when the particle it pertains to encounters a detector the wave function, according to FAPP, collapses and the particle is located at some point in space. In Bohmian mechanics this particle has and always has had a trajectory. What then accounts for equal likelyhood that particles emitted one after the other can land with equal probability in any direction? If we trace the trajectories back to their initial conditions we see that which trajectory we are on depends on the initial values of the wave function. If these are distributed statistically then so will the trajectories. While it is true in Bohmian mechanics that each trajectory is perfectly deterministic, which one we are on is statistically determined.

People have raised various complaints about Bohmian mechanics. A frequent one is that the game of mathematical complexity is not worth the candle of the determinstic interpretation. I do not mind the mathematical complexity so long as there is someone else willing to do the mathematics. A more interesting complaint has to do with the "non-locality" of the theory. This only begins to manifest itself when there is more than one particle involved. Let us consider two particles which are interacted on with a common potential V(**x₁**,**x₂**)**.** The solution to the Schrödinger equation is a function ψ(**x₁**,**x₂**,t). There is a common time because we are dealing non-relativistically. When we use this wave function in the equations of motion of the two particles the density ρ is common but the current **J** is different for each particle since the gradient that defines it is taken with respect to a different variable. Once the particles are in interaction the wave function cannot be written as a product of the wave functions of the individual particles. The particles have become "entangled." This means that determining the trajectory of one particle requires instantanous input from the other no matter how far apart the particles are. It is in this sense that the theory is non-local. If Bohmian mechanics is to agree with ordinary quantum mechanics this kind of non-locality is to be expected. It is well-known that entangled particles produce instantaneous effects on correlated measurements. This is what Einstein referred to as "spooky actions at a distance." No one claims that this is a violation of the theory of relativity here, and there is also no violation in Bohmian mechanics. No information bearing signals are exchanged superluminally.

Bohmian mechanics and the decoherent history interpretation have certain commonalities the most significant of which is that neither accepts the notion that measurements differ in any way from any other kind of interaction. There is no collapse of the wave function. As formulated by Hartle a "history" is a series of answers to "yes"-"no" questions. Is the Moon at such and such a place in its orbit or isn't it? This is a question we can ask over and over again in the course of time and construct a history. To represent this mathematically we correlate to each yes answer a projection operator and to each  no answer an othogonal projection operator. These projection operators evolve in time according to the Heisenberg equation

$$P(t) - \exp(iHt) P(0) \exp(-iHt)$$

To construct a "history" we let a product of these projection operators act at a sequence of times on an initial state $|\psi\rangle$. Following Hartle[7] let us call this sequence of projection operators $C_\alpha$. We make this subscript distinction because there are of course many other possible histories. How probable is this one? By assumption-this is not proved in the many histories interpretation any more than the Born rule is proved in ordinary quantum mechanics-the probabiity $p(\alpha)$ is given by

$$p(\alpha) = \|C_\alpha |\psi\rangle\|^2.$$

Here we must be careful. Not every history has a well defined probability. There can be quantum mechanical interference between histories. In the two slit experiment for example to get the correct probability for the electron to reach a place on the detector you must take the square of the sum of the probability amplitudes and not the sum of the squares. In Bohmian mechanics this issue does not arise since each electron that passes through a slit has a classical

---

[7] See for example James B. Hartle, Quantum Pasts and the Utility of History arXiv:gr-gc/9712001 vl 2 Dec 1997

trajectory. The histories to which one can attach a probability are "decoherent." This means that $\langle\psi|C_\beta^\dagger C_\alpha|\psi\rangle$ is approximately zero onless $\alpha=\beta$. Since we do not require that it be exactly zero this kind of history is "quasi-classical." In Bohmian mechanics the trajectories are really classical. It is argued that for macroscopic objects-the Moon for example-decoherence occurs because of all the collisions between the object and .say. the microwave radiation left over from the Big Bang. In the many histories interpretation what indeed is history?

                      At first sight this might seem to be obvious. All we have to do is to run the chain backwards. Yes this gives one history but there are others, possibly very many others. The reason is that if all we know is the present state vector there are many paths by which we could have arrived there depending on which initial state vector we started from. We have no way of knowing this from the data we have at hand. Let us take an example discussed by Hartle-the Schrödinger cat. I can't resist noting that when I spent an afternoon with Schrödinger in his aparment in Vienna there was no cat. In any event this unfortunate feline is put in a box that contains a capsule of poison gas and a sample of uranium. The capsule is triggered to that if the

uranium has an alpha decay, the alpha sets off the trigger and the unfortunate feline expires. After a time interval we open the box and happily the cat is alive. It could, according to the many history approach have arrived at this state in two ways. The initial state might have been a cat alive state or it might have been a coherent sum of a cat alive and a cat dead state. From the presence of the living can we cannot decide.The vision of the past given by the decoherent history interpretation and the Bohmian seems radically different. In Bohmian mechanics we could in principle follow all the cat molecules backwards in time and arrive at one and only one past.

I don't know how you feel, but the ambiguity of the past makes me queasy. It might be entertaining to imagine that in an alternate past my grandmother who was born in a Polish stetl
could have been Eleanor Roosevelt. I readily accept that these pasts to not communicate but there seem to be too many of them from the point of view of economy. A trip to a barber wielding Occam's razor seems warranted. In any case when it comes to quantum pasts , as Duke Ellington taught us, " Things
ain't what they used to be."